\documentclass[conference]{IEEEtran}
\IEEEoverridecommandlockouts
\usepackage{cite}
\usepackage{amsmath,amssymb,amsfonts}
\usepackage{algorithm} 
\usepackage{algpseudocode}
\usepackage{booktabs}
\usepackage{graphicx}
\usepackage{textcomp}
\usepackage{times}

\def\BibTeX{{\rm B\kern-.05em{\sc i\kern-.025em b}\kern-.08em
    T\kern-.1667em\lower.7ex\hbox{E}\kern-.125emX}}

\begin{document}

\title{Representing Social Networks as Dynamic Heterogeneous Graphs}
\author{\IEEEauthorblockN{Negar Maleki}
\IEEEauthorblockA{\textit{Muma College of Business} \\
\textit{University of South Florida}\\
negarmaleki@usf.edu}
\and
\IEEEauthorblockN{Balaji Padmanabhan}
\IEEEauthorblockA{\textit{Muma College of Business} \\
\textit{University of South Florida}\\
bp@usf.edu}
\and
\IEEEauthorblockN{Kaushik Dutta}
\IEEEauthorblockA{\textit{Muma College of Business} \\
\textit{University of South Florida}\\
duttak@usf.edu}
}

\maketitle

\begin{abstract}
Graph representations for real-world social networks in the past have missed two important elements: the multiplexity of connections as well as representing time. To this end, in this paper, we present a new dynamic heterogeneous graph representation for social networks which includes time in every single component of the graph, i.e., nodes and edges, each of different types that captures heterogeneity. We illustrate the power of this representation by presenting four time-dependent queries and deep learning problems that cannot easily be handled in conventional homogeneous graph representations commonly used. As a proof of concept we present a detailed representation of a new social media platform (Steemit), which we use to illustrate both the dynamic querying capability as well as prediction tasks using graph neural networks (GNNs). The results illustrate the power of the dynamic heterogeneous graph representation to model social networks. Given that this is a relatively understudied area we also illustrate opportunities for future work in query optimization as well as new dynamic prediction tasks on heterogeneous graph structures.
\end{abstract}

\begin{IEEEkeywords}
Heterogeneous Graph, Temporal Graph, GNNs, Steemit
\end{IEEEkeywords}

\section{Introduction}
Social media has become part and parcel of everyday life of individuals. Applications such as Facebook, Snapchat, TikTok, YouTube, LinkedIn, Instagram, Twitter and Reddit have proliferated worldwide; recent statistics suggest that over 10 billion hours a day is spent worldwide on social media alone. As users become more dependent on these platforms, their requirements naturally increase, and platforms need to continue to improve design and features to satisfy new needs. 

As an example, querying social media to determine which users an individual may have interacted with in the last two years is a useful feature to help users maintain their social media accounts in a manner that reflects their own recent interests. Likewise, being able to retrieve posts from specific users within a time frame can help users go back to content they may have recently seen which might be important for some current task. These types of functionalities are still very limited in many social media platforms. Better \emph{representation} of social media data can help platforms offer such functionalities. This paper offers one approach that can help platforms in this process.

Representing social media data as graph network~\cite{1} is a common practice for analytical and machine learning purposes. Graphs, typically, shown as {$G = (V, E)$}, consisting of nodes {$V$} and edges {$E$}. Each node and edge may have its own set of node and edge properties. For example, nodes may represent users and edges may represent the connections between users. Nodes and edges in the graphs will have attributes associated with entities in social media. To fully represent various entities in a social media such as `users', `posts' and `comments' using graphs, a heterogeneous graph structure is needed~\cite{1}. The different relationships among these entities are also depicted using different types of edges such as `authored by', `liked by' and `commented by'. Hence, a representation that can allow both heterogeneity in types of nodes, and heterogeneity in types of edges is important.

Social media data is not static, it is dynamic, and time plays a significant role in how users obtain value from these platforms. Thus, adding a time dimension to a heterogeneous social media graph will provide a representation that permits more analysis on social media data and functionalities in the platforms. However, the time component will add additional complexity to the heterogeneous graph representing the social media data. The time component can help to create a sub-graph out of the full graph that can be used for focused analytical and machine learning tasks on social media data. Still, the heterogeneous graph representing the full dynamic temporal aspect of social media is largely absent in the existing research. In this research, we present a dynamic heterogeneous graph to represent an upcoming social media, Steemit~\cite{2}. We then demonstrate the useful of this representation through implementation of common social media query and predictive analytic task.

A very recent paper~\cite{3} has proposed a dynamic heterogeneous graph with temporal dimension for the similar purpose as described above. Authors of \cite{3} implemented a graph neural network based on heterogeneous temporal graph. In this, an individual heterogeneous graph is created at each $t$ and the relationship between the two graphs in consecutive time $t$ and $(t+1)$ is depicted through additional edges. Thus, if a node and an edge exist in both $t$ and $(t+1)$, the node and the edge will be created twice  - once for graph $G_t$ at $t$ and once for graph $G_{(t+1)}$ at $(t+1)$. Additional edges will be created between nodes and edges in $G_t$ and $G_{(t+1)}$. In a very dynamic situation, such as in social media, where new nodes (depicting post, comment or an user) are added quite frequently, the approach in \cite{3} will create a computationally unmanageable multiple heterogeneous graphs for each time-interval. The representation in \cite{3} will increase the effective number of nodes and edges in the graph. This will increase both storage and computational complexity of using this representation in the context of social media data. Contrary to this approach, in our representation, we embed the time information in the nodes and edges itself without replicating the same node and edges multiple times. This makes the size of the graph manageable, and thus, the approach is scalable. Our representation of dynamic heterogeneous graph is efficient. 

To show the expressive power of our proposed graph representation, we use a social media platform, Steemit \cite{2}, in form of a case study. Steemit is a social media platform that runs on top of a blockchain called Steem, which rewards its users through a cryptocurrency, called Steem dollar. Steemit allows its members to publish content in the form of posts. Posts similar to Reddit are blogs users create and post to the Steemit website. Users can also up-vote/ down-vote other users’ posts, comment on posts, and resteem other users’ posts, which is similar to what retweeting a post on Twitter is or sharing someone else’s post on Facebook. Users can get paid for up-voting/ down-voting and commenting on other posts besides posting a content.

In this paper, we first represent the Steemit data using the proposed dynamic heterogeneous graph structure. Next, we demonstrate how common time dependent queries on social media can be efficiently answered using the proposed representation. Lastly, to demonstrate how GNN models can be adapted for our proposed graph representation, we implement two GNN based prediction model and report their accuracy measures. 

\section{Related Work}
\subsection{Graph Neural Networks}
Graph data represent unique challenges for machine learning algorithms. Graph Neural Networks (GNNs) were developed for such data and have many designs dependent on how these define the concept of a message and the concept of aggregation (i.e. a node's attributes is updated based on messages passed from neighbors). 
Graph Convolutional Network (GCN) \cite{4} is a well-known GNN method that, in terms of weight sharing, is comparable to Convolutional Neural Network (CNN). Fast approximation spectral-based graph convolutional network \cite{5} is one of the key algorithms classified as GCNs and utilized for node clustering. Kipf and Welling \cite{5} built their approach by including the graph's adjacency matrix in the forward propagation equation. 

A limitation in the GCN algorithm is that it assigns the same level of importance to each edge. However, this problem is addressed in Veličković et al. \cite{6} by expanding the aggregation function of the GCN layer by assigning attention coefficients to each edge. The aggregation function in the graph attention network (GAT) is similar to the one in GCN, except neighbors’ embeddings scaled by attention coefficients aggregate together. Algorithms that use low-dimensional node embeddings are used in different prediction tasks. However, most of these algorithms like DeepWalk \cite{7} are inherently transductive; adding a new node requires re-training the model, and cannot be generalized to unseen or new nodes. GraphSAGE is an algorithm that addresses this issue and can be used in dynamic graphs, where the structure is not fixed. GraphSAGE was first introduced in Hamilton et al. \cite{8} to aggregate a new node nearby to generate node embedding for it efficiently. Therefore, we use GraphSAGE in our GNN architecture as the time component makes the graph dynamic.

\subsection{Temporal Graphs}
Learning on temporal graphs is a relatively new field. Temporal graphs was introduced in Kostakos \cite{9} which represented a graph structure including time. The dynamic features of a dataset as well as the entities and connections that it represents can be understood using temporal graphs. 

Most approaches focus on discrete-time dynamic graphs represented as a series of snapshots of the graph \cite{10, 11, 12, 13}; few techniques  support the continuous-time situation have recently been suggested. The Dynamic Graph Neural Network (DGNN) \cite{14} model, used for link prediction and node classification, continuously updates node information by sequentially recording new edges (connections), the time window between edges, and information propagation. The continuous-time dynamic network embeddings (CTDNE) \cite{15} algorithm learns embeddings based on the temporal random walks concept, which is used for link prediction. A temporal walk is a temporally valid sequence of edges walked in ascending order of edge timings that respects time. The system is broadly applicable, with many interchangeable components, and it can be used effectively to include temporal dependencies into current node embedding and deep network models that employ random walks. DyRep \cite{16} is a novel modeling framework that learns richer node representation over time, which is used for link and time prediction. The Temporal Graph Attention (TGAT) \cite{17} layer aggregates temporal and topological features besides learning time-feature interactions. By adding TGAT layers, the network learns node embeddings as a function of time and then can be used for node classification and link prediction tasks. Temporal Graph Network (TGN) \cite{18} framework was developed at Twitter and can be applied to different problems represented as sequences of timed events. 

All the above models are applicable on the homogeneous graph, where there are one type of node and one type of edge. In a recent paper \cite{3}, spatial and temporal dependencies of a heterogeneous graph were integrated to learn node representations over a Heterogeneous Temporal Graph (HTG) - described as an ordered list of heterogeneous graph slices connected by a set of temporal connections. They addressed the similar problem as we do in this paper (i.e. modelling heterogeneous graph with temporal dimensions), but with the difference that we have time in both edges and nodes of our heterogeneous graph instead of having them only for slicing the graph. To this end, like \cite{3}, we use time in slicing graph for train-validation-test sets. Moreover, unlike previous graph representations, we include time in every component of the graph, i.e, nodes and edges and use them in GNN.

\subsection{Multiplex Network}
Multiplexity refers to multifaceted connections between two persons \cite{19}; therefore, social networks are a good example of multiplex networks. A group of networks containing different types of relations is considered a multiplex network where each kind of relationship creates a layer of the network.

Real-world networks are significantly sophisticated, as cross-domain interactions across various networks are well-observed, resulting in a sort of multi-layered network \cite{20}. A recent paper studied how to perform network embedding for nodes on multiplex networks by providing a unified optimization framework, called MANE \cite{20}. Their framework is considered as a heterogeneous information network, since different node types are placed in different layers. However, having different layers with different node types adds to the complexity of multiplex networks. Thus, researchers focus on different types of layers with a single node type and suggest a multiplex network embedding model, called MNE \cite{21}. However, this framework aggregates information over all layers even the irrelevant ones, so it is not computationally efficient. To address this drawback, the DEEPLEX \cite{22} framework was proposed which considers sampling k-nearest layers with the most similar embeddings. In a recent paper \cite{23}, a framework is suggested to be more efficient in terms of computational complexity which addressed the MNE problem by selecting information from relevant layers, and also addressed DEEPLEX drawback by adaptively learning a sampling distribution over the relevant layers. The idea of having information from relevant layers inspired us to extract sub-graphs out of the larger heterogeneous graph using time component queries. Although extracting sub-graphs decreased embedding and topological information, our findings show that we still get at least the same or better results than including all information in the payout classification problem, besides the obvious time efficiency.

\section{Graph Representation}
A graph is a generic mathematical language for describing complex networks. Graphs are commonly represented as {$G = (V, E)$}, with nodes {$V$} and edges {$E$}. In this paper, the graph is constituted of actors and actionable items {$< actor\textsubscript{1} - action \rightarrow actor\textsubscript{2} >$}, where each may have its own set of attributes. Both the actor and the actionable item have time attribute associated with it. For actors, the time indicates when the actor was created (such as `joined' for user and `created' for post and comments). The actionable item has time attribute indicating when the action (such as authored or voted) took place.  

Attributes can be divided into two main categories, static and dynamic attributes. Static attributes mainly define the state of an actor or actionable item (such as `permlink' and `category'). However, dynamic attributes as their name states can be changed over time, so they are time-dependent attributes (such as the `payout', `author\_reputation', `author\_rewards'). Networks, in reality, are not static, which means actors take actions and modify the networks in various ways. 

Three main operations take place in networks: items addition, items deletion, and items modification. In items addition operation, new actors or actionable items appear and create new connections to the network, which extend the network. Items modification operation gives actors or actionable items an opportunity to update or modify the existing actors or actionable items. Lastly, the items deletion operation removes existing actors or actionable items where all the dependent children will be disappeared as their parents are removed.

Graphs are searchable and able us to query and extract intuitive information, which later can be used in different ML tasks. Relational algebra operations like selection, projection, and join are common among different databases management systems (DBMS), including network database systems. Selection ({$\sigma$}) operation as one of the unary relational operations is used for selecting a subset of tuples that satisfy the selection condition. Projection ({$\pi$}) operation as another operation in the unary relational operations is used for keeping specific attributes from a relation and ignoring the rest attributes. Lastly, join ({$\bowtie$}) operation as one of the binary relational operations is used for joining variously related tuples from different relations.

We followed the representation in \cite{1} to illustrate Steemit in form of a graph in  Fig~\ref{steemit-graph}. However, we have adapted it to the actions in Steemit (such as \emph{vote} vs. \emph{like}) and added the time attribute in the graph in Fig~\ref{steemit-graph}. 

\begin{table}[htbp]
  \centering
  \scriptsize
  \caption{Actor and actionable items types}
  \label{edge-action}
    \begin{tabular}{lll}
    \toprule
    \textbf{Item}  & \textbf{Item Type} & \textbf{Description} \\
    \toprule
    authored & Actionable & A person is the author of a post. \\
    \midrule
    reply & Actionable & A post is replied by a comment. \\
    \midrule
    vote  & Actionable & A post is voted by a person. \\
    \midrule
    follow & Actionable & A person is followed by a person. \\
    \midrule
    user  & Actor & Take actionable items on another actor. \\
    \midrule
    post  & Actor & Take actionable items on another actor. \\
    \midrule
    comment & Actor & Take actionable items on another actor. \\
    \bottomrule
    \end{tabular}%
\end{table}%

\subsection{Dataset}
For demonstration purposes we use the SteeemOPS data from Steemit in this paper. SteemOps is a dataset presented in \cite{24}, which contains ten key types of Steemit operations organized into three sub-datasets: (1) the social-network operation dataset (SOD), (2) the witness-election operation dataset (WOD), and (3) the value-transfer operation dataset (VOD). The data was collected from 2016/03/24 16:05:00 to 2019/12/01 00:00:00. The main sub-dataset we use in this paper is SOD consisting of three operational keys - comment, vote, and custom-json. The comment operation consists of five fields, see Table~\ref{table:6}. According to \cite{24}, a new post is indicated when both parent-author and parent-permlink fields are empty. When these two fields are not empty, it represents a comment to a post/comment. Each post in Steemit remains active for 7-day, so each time the author made any changes to his post, the post will be recorded as a new post to the dataset. Moreover, each post's permlink is unique, so considering all of these, the dataset consists of 17805355 new posts.

\begin{table}[htbp]
  \caption{Schema of operation comment \cite{24}}
  \label{table:6} 
  \resizebox{\columnwidth}{!}{%
    \begin{tabular}{ccl}
    \toprule
    \multicolumn{1}{c}{\textbf{Field name}} & \multicolumn{1}{c}{\textbf{Type}} & \multicolumn{1}{l}{\textbf{Description}} \\
    \midrule
    block\_no & Integer & the block recording this operation \\
    \midrule
    parent\_author  & String  & the author that comment is being submitted to \\
    \midrule
    parent\_permlink & String  & specific post that comment is being submitted to \\
    \midrule
    author & String  & author of the post/comment being submitted (account name) \\
    \midrule
    permlink  & String  & unique string identifier for the post, linked to the author of the post \\
    \bottomrule
    \end{tabular}%
    }
\end{table}%

The Steem platform offers an interactive Application Programming Interface (API) for researchers to parse the data. However, just retrieving the full information considering some API restrictions would have taken approximately 38 days in total, so we retrieved a random 10\% of this dataset (approximately 1.7 million new posts) for demonstration purposes in this paper. For these 1.7 million posts, we also retrieved relative information like comments of the posts and voters. We randomly selected half of August 2019 data and converted it into graph to demonstrate in this paper. The entire heterogeneous graph contains 165180 nodes and 1405939 edges after construction, which covers 24000 posts published in August 2019.

\subsection{Heterogeneous Graph Construction}\label{label:Heterogeneous Graph Construction}
We represent Steemit as a directed graph {$< V, E >$}, where {$V$} represents actors (nodes), and {$E$} represents actionable items (edges). In Steemit, an actor can be of different types, such as user and content (post or comment). The interactions between actors are captured by actionable items which also can be of different types, see Table~\ref{edge-action}. For example, a user can follow another person, which we denote as $< user\textsubscript{1} - follow \rightarrow user\textsubscript{2} > \in E$. Similarly, a user can make a post, which we denote as $< user\textsubscript{1} - Authored \rightarrow post\textsubscript{1} > \in E$, or a user can vote a post or a comment, which we denote as \{$< user\textsubscript{1} - vote \rightarrow post\textsubscript{1} >$, $< user\textsubscript{1} - vote \rightarrow comment\textsubscript{1} >$\} {$\subset E$}. 

\begin{figure}[htbp]
    \centering
    \includegraphics[width=\columnwidth]{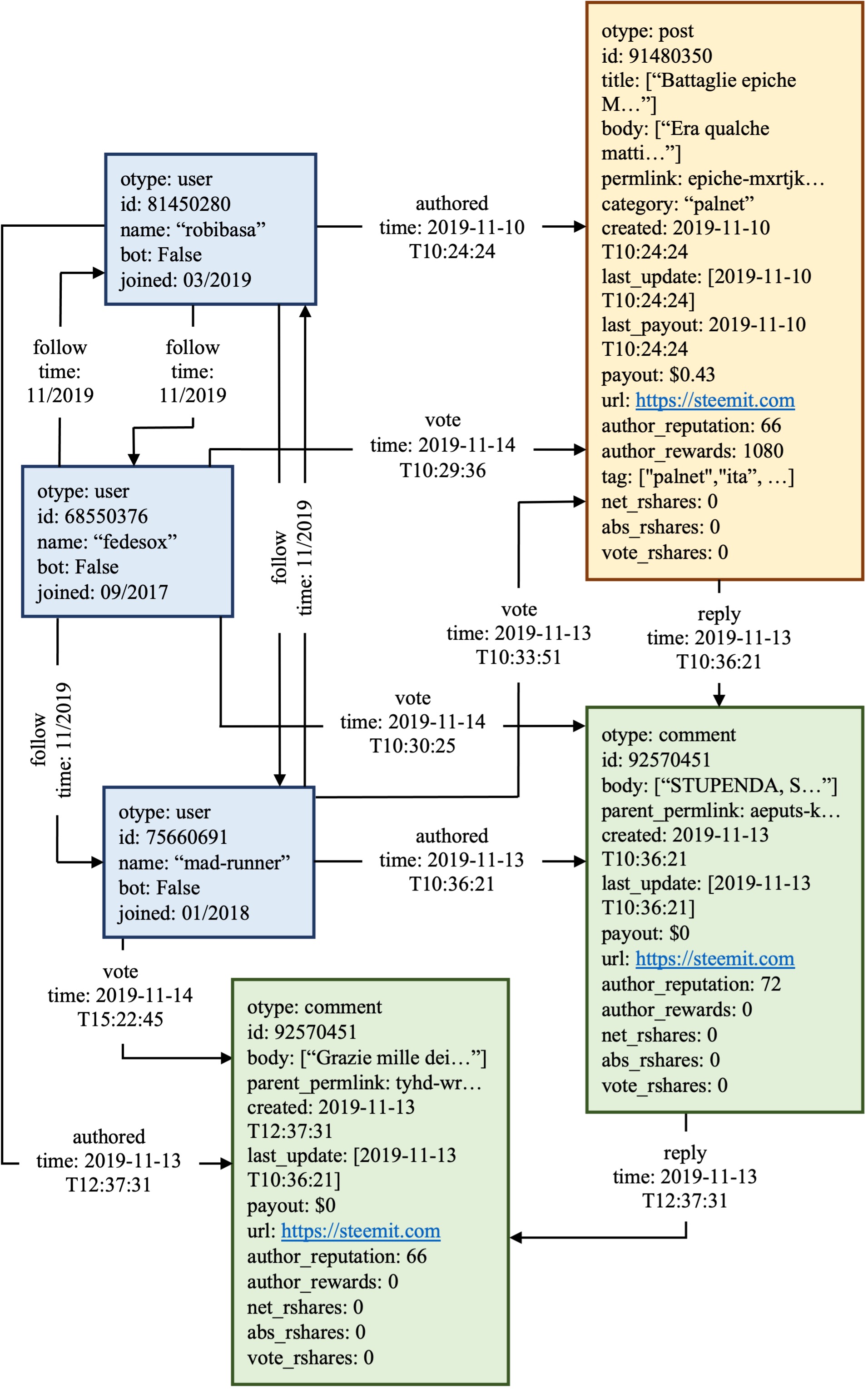}
    \caption{Steemit graph structure}
    \label{steemit-graph}
\end{figure}

The heterogeneity of the graph is indicated through the attribute `otype' in the nodes indicating actors, which can be `user', `comment' or `post'. Similarly the edges indicating actionable items can be of different types 'follow', `vote', `authored' and `reply'. As mentioned before, actors and actionable items have static and/ or dynamic attributes. For example, user object type (otype) or actor has “name”, “bot”, and “joined” date as static attributes and number of followers and following as dynamic attribute that can be calculated using graph queries. The "post" object type (otype) or actor has many static attributes illustrated in Fig.~\ref{steemit-graph}, except “title”, “body”, and “last\_update” which are considered as dynamic attributes where their changes will be stored in form of a list or JSON depending on the attribute characteristics, and also a total number of post’s votes can be calculated using graph queries (sum of incoming votes (actionable item)). Note that the dynamic attributes calculated through queries do not need to be explicit in the graph representation. Both the actions (nodes) and the actionable items (edges) have time as one attribute which makes it a \emph{Dynamic Heterogeneous Graph}. 

We constructed the Steemit graph as a directed graph (DiGraph) using the NetworkX Python library. The constructed graph is then converted into the PyTorch Geometric (PyG) graph in order to be used for GNN models. By doing so, we create a dynamic graph from which needed sub-graphs can be extracted using queries, and the derived sub-graphs may subsequently be utilized in various deep learning tasks.

We demonstrate the utility of this heterogeneous dynamic graph representation of Steemit with the help of common queries and two application of GNN machine learning on this graph representation.

\section{Graph Queries}
The entire graph is not always the point of interest in different tasks to be performed on social networks such as the Steemit graph. The unique aspect of Steemit graph representation is the time attribute in both actions and actionable items. To this end, we may like to take a sub-graph out of the entire graph, which represents all activities between a specific time interval. The application of such query could be many fold; for example, in presidential election seasons, people's activities on social media become attractive. Therefore, analysis of such activities gives interesting information required to be gained from specific time intervals. The key point in this representation is that we only include actions and actionable items relevant in the specified time intervals. Any actions or actionable items connected to nodes but did not occur during the given time interval will be removed from the sub-graph. Algorithm ~\ref{query1} shows the pseudo code of the sub-graph query between {$t\textsubscript{1}$} and {$t\textsubscript{2}$}, which is the first portion for all the subsequent queries discussed in this section. In the following queries, the output of the Algorithm~\ref{query1} will be given as an input. The time complexity of the Algorithm~\ref{query1} is {$O(n+m)$} where $n$ is the number of nodes and $m$ is the number of edges. 

\begin{algorithm}[htbp]
	\caption{Query Sub-Graph between {$t\textsubscript{1}$} and {$t\textsubscript{2}$}} 
	\label{query1}
	\small
	\begin{algorithmic}[1]
	\State \textbf{Input:} {$G (V, E)$}: graph with {$n$} nodes and {$m$} edges
    \State \textbf{Input:} t\textsubscript{1}, t\textsubscript{2}: timestamp
    \State \textbf{Output:} {$G\_post\_t\textsubscript{1}\_t\textsubscript{2} (A, B)$} : Subgraph between {$t\textsubscript{1}$} and {$t\textsubscript{2}$}
    \State \textbf{define function} get\_key(val):
    \State \textbf{set} key\_indx\_node \textbf{to} []
    \State \textbf{set} key\_indx\_edge \textbf{to} []
	\For{{$key, value$} \textbf{in} networkX.get\_attributes({$G$}, 'otype').items()}
	\If {val \textbf{equals} value} 
	\State \textbf{add} key \textbf{to} key\_indx\_node
	\State \textbf{add} key \textbf{to} key\_indx\_edge
	\EndIf
	\EndFor
    \State \textbf{return} key\_indx\_node, key\_indx\_edge
	\State \textbf{call set} post, comment, user \textbf{to} get\_key(['post', 'comment', 'user'])
	\State \textbf{call set} vote, authored, reply \textbf{to} get\_key(['vote', 'authored', 'reply'])
	\State \textbf{set} subG\_post, subG\_comment \textbf{to} []
    \For{i \textbf{in} post (or comment)}
    \If{{$t\textsubscript{1} \leq$} G.nodes()[i]['created'] {$\leq t\textsubscript{2}$}}
    \State \textbf{add} i \textbf{to} subG\_post (or subG\_comment)
    \EndIf
    \EndFor
	\State \textbf{set} subG\_vote, subG\_authored, subG\_reply \textbf{to} []
    \For{i \textbf{in} vote (or authored or reply)}
    \If{{$t\textsubscript{1} \leq$} G.edges()[i]['time'] {$\leq t\textsubscript{2}$}}
    \State \textbf{add} i \textbf{to} subG\_vote (or subG\_authored or subG\_reply)
    \EndIf
    \EndFor
    \State \textbf{set} G\_node \textbf{to} G.subgraph(subG\_post + subG\_comment + user)
    \State \textbf{set} G\_post\_t\textsubscript{1}\_t\textsubscript{2} \textbf{to} G\_node.edge\_subgraph(subG\_vote + subG\_authored + subG\_reply)
	\end{algorithmic} 
\end{algorithm}

\begin{figure}[htbp]
    \centering
    \includegraphics[width=\columnwidth]{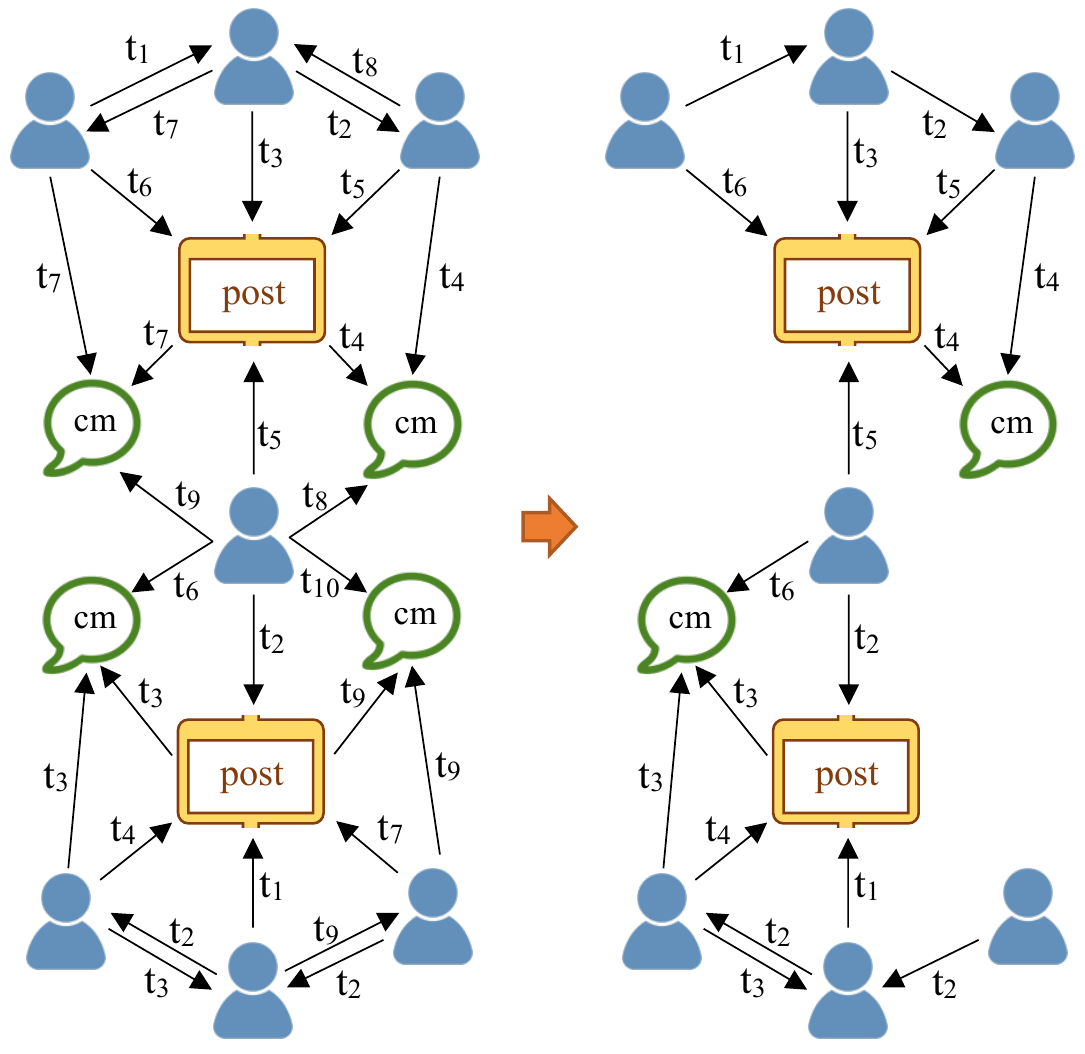}
    \caption{Result of the Algorithm~\ref{query1}: sub-graph between {$t\textsubscript{1}$} and {$t\textsubscript{6}$}}
    \label{Q1}
\end{figure}

Consider Fig.~\ref{Q1}, which shows a heterogeneous network including users, posts, and comments as nodes and edges connecting these nodes (node types and edge types are as described in Table~\ref{edge-action})). Along with each edge, the $t$ shows when the actionable items took place, and the subscripts show the order of those actionable items, such as $t\textsubscript{2}$ $>$ $t\textsubscript{1}$. Fig.~\ref{Q1} shows the result of Algorithm~\ref{query1} if we are interested in constructing the sub-graph between {$t\textsubscript{1}$} and {$t\textsubscript{6}$}. All the nodes and edges that are not placed in the given time interval will be deleted and the final sub-graph will include exactly actions and actionable items between {$t\textsubscript{1}$} and {$t\textsubscript{6}$}.

To show the expressive power of time in the Steemit graph representation, the sub-graph extracted for a given time interval will be used as the input of the following queries. We present three different queries that has common applications in the social media context. 

\subsection{Querying on Category of Posts}
The category of a post is one of the attributes in Steemit. This attribute can be used for selecting posts in specific field. For example, for health-related posts, we may look for categories such as 'healthcare' or 'medicine'.  Algorithm~\ref{query1_1} is the query following the sub-graph between {$t\textsubscript{1}$} and {$t\textsubscript{2}$} query to select posts with a specific category within that time frame. The time complexity of this algorithm is {$O(n^2)$} where $n$ is the number of nodes in sub-graph between {$t\textsubscript{1}$} and {$t\textsubscript{2}$}.

\begin{algorithm}[htbp]
	\caption{Query SubGraph between {$t\textsubscript{1}$} and {$t\textsubscript{2}$} with an specified category} 
	\label{query1_1}
	\small
	\begin{algorithmic}[1]
	\State \textbf{Input:} {$G\_post\_t\textsubscript{1}\_t\textsubscript{2} (A, B)$} : Subgraph between {$t\textsubscript{1}$} and {$t\textsubscript{2}$}
    \State \textbf{Input:} category: string
    \State \textbf{Output:} {$G\_post\_t\textsubscript{1}\_t\textsubscript{2}\_category (C, D)$} : Subgraph between {$t\textsubscript{1}$} and {$t\textsubscript{2}$} with an specified category
    \State \textbf{call set} post\_ \textbf{to} get\_key('post')
    \State \textbf{set} subG\_t\textsubscript{1}\_t\textsubscript{2}\_post \textbf{to} []
    \For{i \textbf{in} post\_}
    \If{G\_post\_t\textsubscript{1}\_t\textsubscript{2}.nodes()[i]['category'] \textbf{equals} category}
    \State \textbf{add} i \textbf{to} subG\_t\textsubscript{1}\_t\textsubscript{2}\_post
	\EndIf
	\EndFor
    \State \textbf{set} related\_node \textbf{to} []
    \State \textbf{set} related\_cm \textbf{to} []
    \For{i \textbf{in} subG\_t\textsubscript{1}\_t\textsubscript{2}\_post}
    \For{j \textbf{in} range(len(subG\_vote))}
    \If{i \textbf{equals} subG\_vote[j][1]}
    \State \textbf{add} subG\_vote[j][0] related\_node
    \For{j \textbf{in} range(len(subG\_authored))}
    \If{i \textbf{equals} subG\_authored[j][1]}
    \State \textbf{add} subG\_authored[j][0] related\_node
    \For{j \textbf{in} range(len(subG\_reply))}
    \If{i \textbf{equals} subG\_reply[j][0]}
    \State \textbf{add} subG\_reply[j][1] related\_node
    \State \textbf{add} subG\_reply[j][1] related\_cm
    \EndIf
    \EndFor
    \EndIf
    \EndFor
    \EndIf
    \EndFor
    \EndFor
    \For{i \textbf{in} related\_cm}
    \For{j \textbf{in} range(len(subG\_vote))}
    \If{i \textbf{equals} subG\_vote[j][1]}
    \State \textbf{add} subG\_vote[j][0] related\_node
    \For{j \textbf{in} range(len(subG\_authored))}
    \If{i \textbf{equals} subG\_authored[j][1]}
    \State \textbf{add} subG\_authored[j][0] related\_node
    \EndIf
    \EndFor
    \EndIf
    \EndFor
    \EndFor
    \State \textbf{set} category\_node \textbf{to} drop.duplicates(related\_node)
    \State \textbf{set} G\_post\_t\textsubscript{1}\_t\textsubscript{2}\_category \textbf{to} G\_post\_t\textsubscript{1}\_t\textsubscript{2}.subgraph (category\_node + subG\_t\textsubscript{1}\_t\textsubscript{2}\_post)
	\end{algorithmic} 
\end{algorithm}

\begin{figure}[htbp]
    \centering
    \includegraphics[width=\columnwidth]{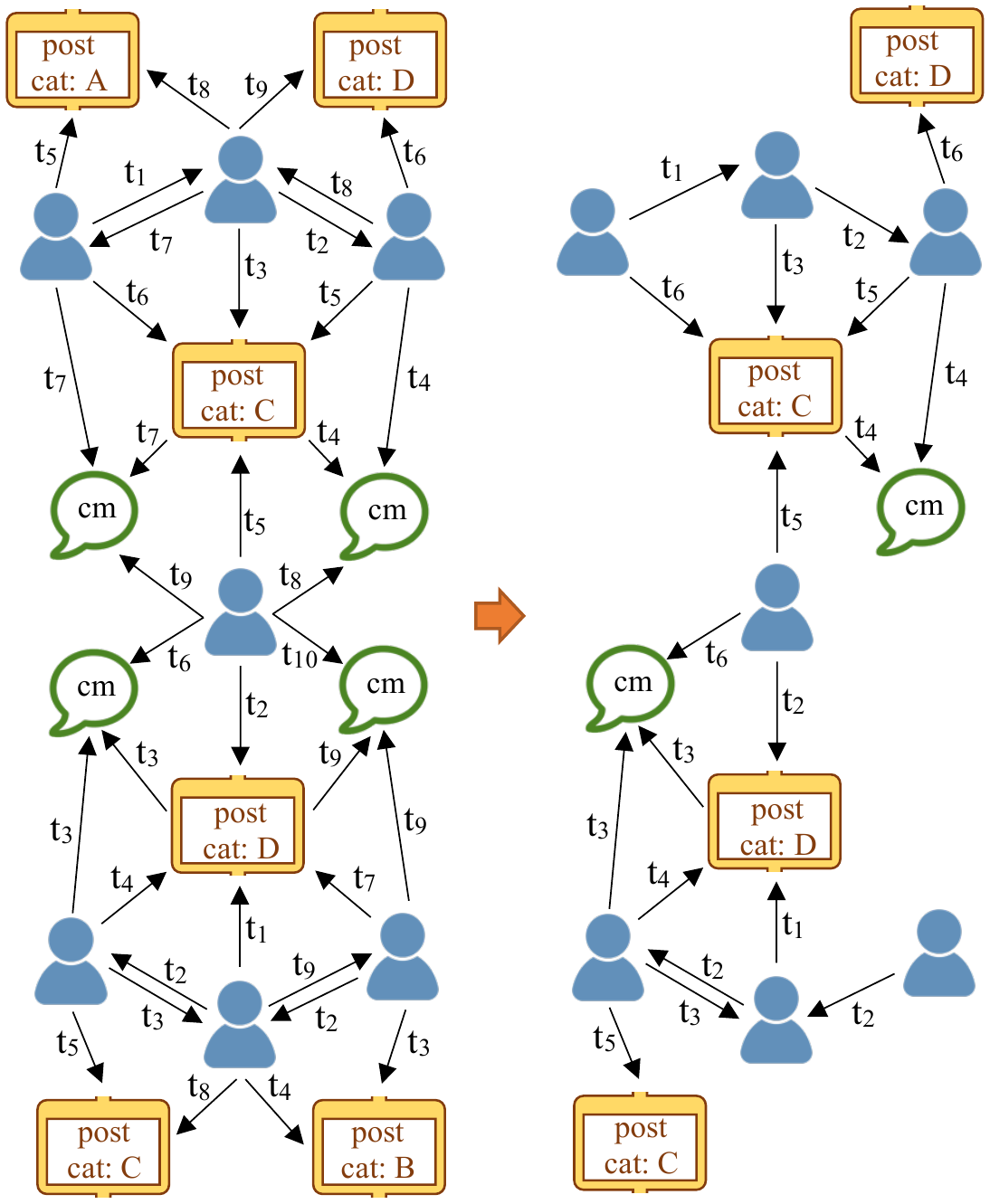}
    \caption{Result of the \textbf{Algorithm} \textbf{\ref{query1_1}}: sub-graph between {$t\textsubscript{1}$} and {$t\textsubscript{6}$} with C and D categories in posts}
    \label{Q2}
\end{figure}

Consider Fig.~\ref{Q2}, which shows a heterogeneous network with nodes representing users, posts, and comments. The goal is to have specific posts in specific category between a given time interval. Posts' categories are {$A, B, C, D$}, and we are interested in {$C$ and $D$} categories between {$t\textsubscript{1}$} and {$t\textsubscript{6}$}. After applying Algorithm~\ref{query1_1} to this graph, any nodes and edges that are not placed in the given time interval and any posts that do not belong to {$C$ and $D$} will be eliminated. Therefore, the final sub-graph will include actions and actionable items between {$t\textsubscript{1}$} and {$t\textsubscript{6}$}, as well as posts with {$C$} and {$D$} categories only.

\subsection{Popularity Query}
The two most important factors for determining how much payout posts get (denoting popularity) in Steemit \cite{24} are the number of votes and comments of a post. Thus to identify the posts that will have the higher payouts (or more popular), we may sort posts based on number of votes and comments Algorithm~\ref{query1_2} is the query on the sub-graph between {$t\textsubscript{1}$} and {$t\textsubscript{2}$} to sort posts based on their vote and comment count. Its time complexity is {$O(nlog(n)+nm)$}, where $n$ is the number of nodes and $m$ is the number of edges in sub-graph between {$t\textsubscript{1}$} and {$t\textsubscript{2}$}. If {$O(nm) \gg O(nlog(n))$}, the time complexity is {$O(nm)$}, or vice versa.

\begin{algorithm}[htbp]
	\caption{Query to Sort posts based on their vote and comment count} 
	\label{query1_2}
	\small
	\begin{algorithmic}[1]
	\State \textbf{Input:} {$G\_post\_t\textsubscript{1}\_t\textsubscript{2} (A, B)$} : Subgraph between {$t\textsubscript{1}$} and {$t\textsubscript{2}$}
    \State \textbf{Output:} comments\_sorted, votes\_sorted: Lists of posts in descending order based on their comment and vote count
    \State \textbf{set} vote\_count \textbf{to} []
    \State \textbf{set} comment\_count \textbf{to} []
    \For{i \textbf{in} {$G\_post\_t\textsubscript{1}\_t\textsubscript{2}$}.nodes()}
    \State \textbf{set} vote \textbf{to} 0
    \State \textbf{set} comment \textbf{to} 0
    \If{{$G\_post\_t\textsubscript{1}\_t\textsubscript{2}$}.nodes()[i]['otype'] \textbf{equals} 'post'}
    \For{j \textbf{in} $G\_post\_t\textsubscript{1}\_t\textsubscript{2}$.out\_edges(i)}
    \If{$G\_post\_t\textsubscript{1}\_t\textsubscript{2}$.edges()[i,j[1]]['otype'] \textbf{equals} 'reply'}
    \State comment += 1
    \EndIf
    \EndFor
    \State \textbf{add} [i, comment] \textbf{to} comment\_count
    \For{j \textbf{in} {$G\_post\_t\textsubscript{1}\_t\textsubscript{2}$}.in\_edges(i)}
    \If{ {$G\_post\_t\textsubscript{1}\_t\textsubscript{2}$}.edges()[j[0], i]['otype'] \textbf{equals} 'vote'}
    \State vote += 1
    \EndIf
    \EndFor
    \State \textbf{add} [i, vote+1] \textbf{to} vote\_count
    \EndIf
    \EndFor
    \State \textbf{set} comments\_sorted \textbf{to} sorted(comment\_count, key=lambda x: x[1], reverse \textbf{to} \textbf{TRUE})
    \State \textbf{set} votes\_sorted \textbf{to} sorted(vote\_count, key=lambda x: x[1], reverse \textbf{to} \textbf{TRUE})
	\end{algorithmic} 
\end{algorithm}

As an example of the larger Steemit graph, consider that we are interested in sorting posts based on their votes and comments count between "Thursday, August 1, 2019 4:00:00 AM GMT" and "Saturday, August 10, 2019 4:00:00 AM" (10-day). In the entire heterogeneous graph, we have 165180 nodes and 1405939 edges. Out of this,  within the specified time interval, the number of nodes and edges are 85062 and 907437, respectively. Table~\ref{result-q3} includes 10 top posts' title and payout with their number of votes and comments within given time interval. Total execution time of this query was 25.20397 seconds; Seconds on a 2.7GHz processor with 16GB RAM.

\begin{table}[htbp]
  \centering
  \caption{Top 10 posts with highest number of comments and votes between "Thursday, August 1, 2019 4:00:00 AM GMT" and "Saturday, August 10, 2019 4:00:00 AM"}
  \label{result-q3} 
  \resizebox{\columnwidth}{!}{%
    \begin{tabular}{p{20.25em}ccc}
    \toprule
    \multicolumn{1}{c}{\textbf{Posts Title}} & \textbf{Payout} & \textbf{Num. of Comments} & \textbf{Num. of Votes} \\
    \midrule
    Announcing the eSteem Token: ESTM & 142.047 & 71    & 555 \\
    \midrule
    IT IS A FLOWER AND BENEFICIAL INSECT WEEKEND! & 3.477 & 66    & 225 \\
    \midrule
    "Guess Who Is In This Photo" Contest (Sport\textbackslash{}'s Figure) & 0.103 & 62    & 116 \\
    \midrule
    BLOODY IMPORTANT QUESTION: should we consider STEEM-ENGINE tokens a security or utility? & 52.223 & 60    & 325 \\
    \midrule
    "Guess Who Is In This Photo" Contest (Sport\textbackslash{}'s Figure) & 0.065 & 57    & 131 \\
    \midrule
    Giving Away 500 Steem To One Random Person & 3.926 & 55    & 136 \\
    \midrule
    "Guess Who Is In This Photo" Contest (Sport\textbackslash{}'s Figure) & 0.093 & 54    & 77 \\
    \midrule
    Actifit Major Announcement: Move AFIT To Steem-Engine. AFITX Our New Exclusive Token Announced - Detailed Dive. Airdrop Announce. Daily Updates & 107.177 & 53    & 673 \\
    \midrule
    Pony Auction on Market Friday & 5.105 & 52    & 159 \\
    \midrule
    "Guess Who Is In This Photo" Contest (Sport\textbackslash{}'s Figure) & 0.035 & 51    & 76 \\
    \bottomrule
    \end{tabular}%
    }
\end{table}%

\subsection{Actively Engaged Query}
Social media platforms quickly became the most used online services, through which billions of users interact intensively every day. This makes social media a significant resource for advertising, marketing, or politics, collecting information, and launching campaigns. A challenging problem is identifying actively engaged users on Steemit (or any social media), where an actively engaged user is described as an user with a higher number of votes, comments, and the number of posts they have published. Algorithm~\ref{query1_3} is the query on the sub-graph between {$t\textsubscript{1}$} and {$t\textsubscript{2}$} to sort users based on their activities within that time interval. Its time complexity is {$O(n)$}, where $n$ is the number of nodes in sub-graph between {$t\textsubscript{1}$} and {$t\textsubscript{2}$}.

\begin{algorithm}[htbp]
	\caption{Query to Sort users based on Activities} 
	\label{query1_3}
	\small
	\begin{algorithmic}[1]
	\State \textbf{Input:} {$G\_post\_t\textsubscript{1}\_t\textsubscript{2} (A, B)$} : Subgraph between {$t\textsubscript{1}$} and {$t\textsubscript{2}$}
    \State \textbf{Output:} user\_activity\_sorted: List of users in descending order who are actively engaged
    \State \textbf{set} user\_activity \textbf{to} []
    \For{i \textbf{in} range(len(user))}
    \If{user[i] \textbf{in} {$G\_post\_t\textsubscript{1}\_t\textsubscript{2}$}.nodes()}
    \State \textbf{add} [user[i], len({$G\_post\_t\textsubscript{1}\_t\textsubscript{2}$}.out\_edges(user[i]))] \textbf{to} user\_activity
    \EndIf
    \EndFor
    \State \textbf{set} user\_activity\_sorted \textbf{to} sorted(user\_activity, key=lambda x: x[1], reverse \textbf{to} \textbf{TRUE})
	\end{algorithmic} 
\end{algorithm}

As another example, consider we are interested in knowing users who are actively engaged in Steemit platform between "Thursday, August 1, 2019 4:00:00 AM GMT" and "Saturday, August 10, 2019 4:00:00 AM" (10-day). Table~\ref{result-q4} includes 10 top users with the number of activities in casting votes and writing posts or comments. Total execution time of this query was 14.1105 seconds; Seconds on a 2.7GHz processor with 16GB RAM. (Note that in Steemit usernames are anonymous but publicly available.)

\begin{table}[htbp]
  \centering
  \caption{Top 10 active engaged users between "Thursday, August 1, 2019 4:00:00 AM GMT" and "Saturday, August 10, 2019 4:00:00 AM"}
  \label{result-q4} 
  \resizebox{\columnwidth}{!}{%
    \begin{tabular}{ccc}
    \toprule
    \textbf{Username} & \textbf{Num. of Cast Votes} & \textbf{Num. of Written Posts/Comments} \\
    \midrule
    'laissez-faire' & 5352  & 0 \\
    \midrule
    'steemitboard' & 1235  & 422 \\
    \midrule
    'accelerator' & 2547  & 5 \\
    \midrule
    anomaly' & 2459  & 0 \\
    \midrule
    'imisstheoldkanye' & 2444  & 0 \\
    \midrule
    'actifit' & 994   & 192 \\
    \midrule
    'map10k' & 1730  & 2 \\
    \midrule
    'fyrstikken' & 1739  & 0 \\
    \midrule
    'hdu' & 1513  & 0 \\
    \midrule
    'steem-plus' & 686   & 134 \\
    \bottomrule
    \end{tabular}%
    }
\end{table}%

\subsection{Query Performance}
\begin{figure}[htbp]
    \centering
    \includegraphics[width=\columnwidth]{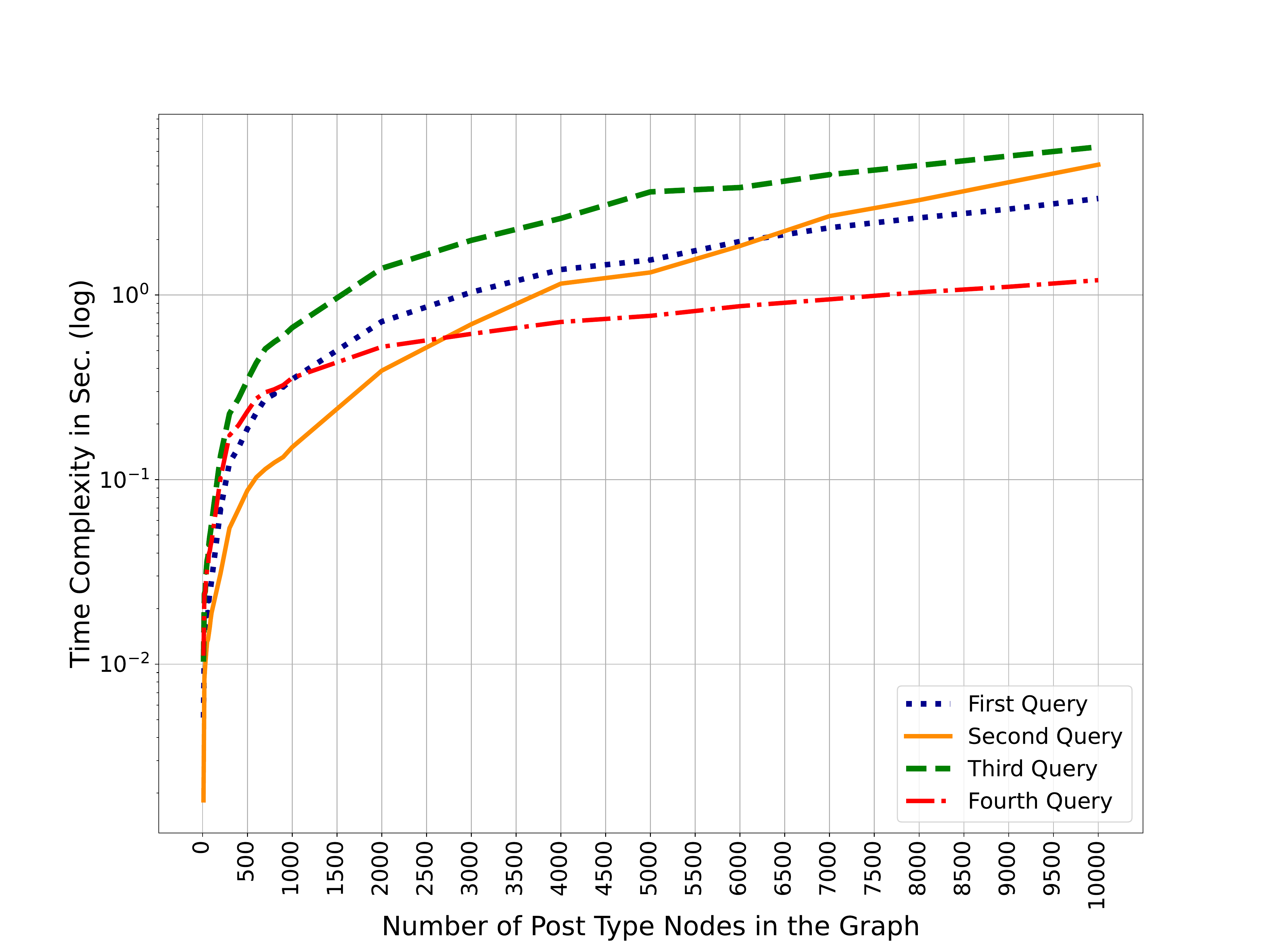}
    \caption{Query time complexity}
    \label{time-complexity}
\end{figure}

As we discussed in the previous sub-sections, these queries have a low execution time compared to the size of the graph (some examples illustrated run times on a single machine; here we present query performance results on a compute cluster). Fig.~\ref{time-complexity} illustrates the execution time of each query for different number of initial set of posts on 1 node with 128 AMD Epyc 7702 CPU @ 2.0GHz processors and 1 TB memory. We focused on the number of posts in each run and implemented four queries for those posts. We considered the same $t\textsubscript{1}$ and $t\textsubscript{2}$ for all the queries. For the second query, we retrieved posts with the "Steemit" category. As demonstrated in the Fig.~\ref{time-complexity}, the query execution time even for 10K posts is below 10 seconds. However, we note that the focus in this paper is a new representation for social network data that combines dynamic and heterogeneous graphs, and showing the value of this representation through illustrative dynamic queries and prediction tasks. We have not addressed the issues of query optimization for real-time performance, which is an important topic for future work.

\section{Illustrative Prediction Tasks}
In this section we show how the dynamic heterogeneous network representation can also be used in illustrative prediction tasks. Predicting the payout for a newly published post is one of the most important classification problems in the Steemit setting. Being able to predict the financial reward from a post can be used in ranking algorithms for content and can also be used as a tool to help users develop better content to post on social media. This problem is also equivalent to predicting whether a post in traditional social media will gain popularity in terms of number of comments and likes received. 

\begin{figure}[htbp]
    \centering
    \includegraphics[width=\columnwidth]{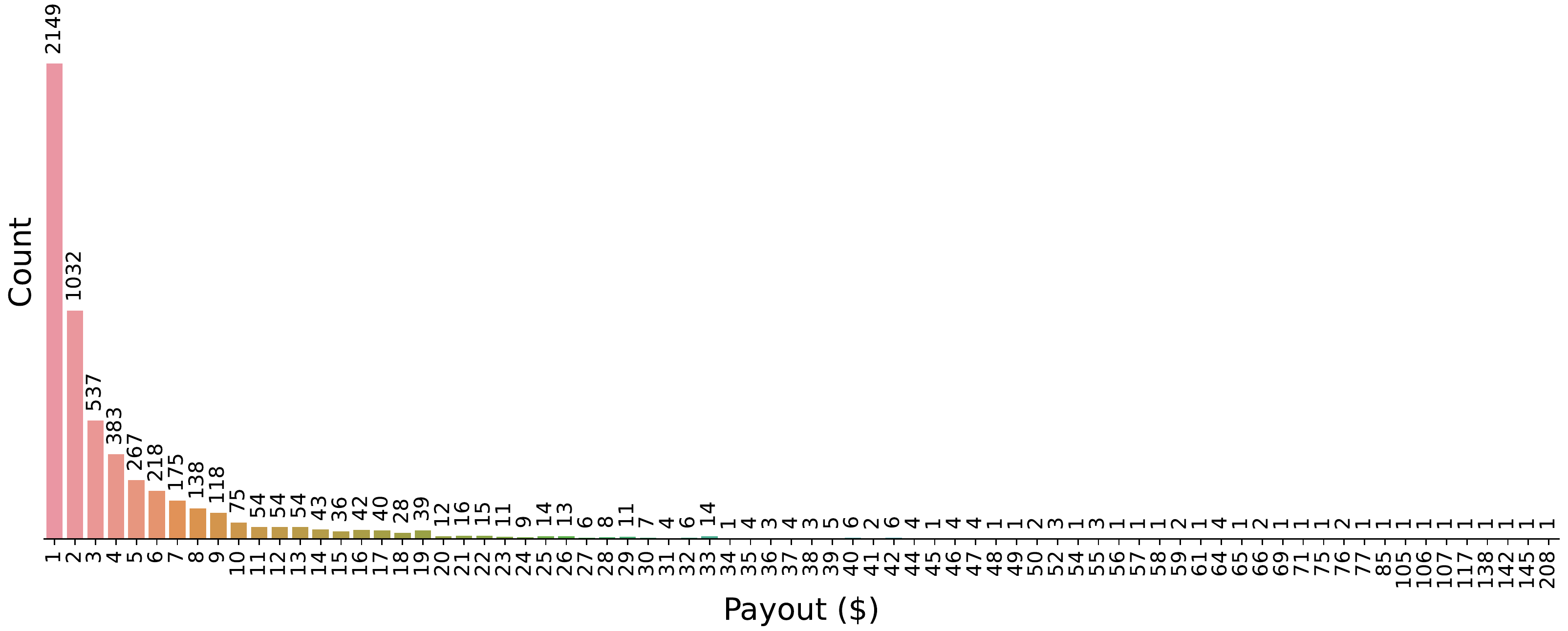}
    \caption{Posts' payout distribution}
    \label{payout-dist}
\end{figure}

According to the Steemit documentation, the payout is determined by a variety of factors, including the number of up-votes and down-votes, the number of comments, the reputation scores of users who voted or replied to the post, how much new Steem Dollar (the cryptocurrency in Steemit platform) is created each week and distributed to users as rewards over a rolling 7-day period. As there are so many dynamic elements that impact the ultimate payout, no exact formula defines what the payout of a new post will be. As shown in Table~\ref{result-q3}, posts with high number of comments and votes did not necessarily get paid highly. Although predicting payout seems to be a regression task, the unavailability of time series data on payouts limited us to defining this problem as a classification task. We have defined this problem as a classification problem that classified the payout amount of a Steemit post into low, medium, and high categories. Fig.~\ref{payout-dist} shows the distribution of number of posts with non-zero payout. Here, we use this prediction problem to illustrate the kind of deep learning approach can be applied in our dynamic heterogeneous graph representation of social network data.

\subsection{Predicting Payout of a Post using Post Features}\label{label:post-feature}
In this approach, we include all the published features in the dataset (see Table~\ref{features}) for each type of node and edge, along with topological information of the graph, and then predict the payout of a new post. As we have time in each component of the graph, we ensure using queries discussed earlier that all the connections to each post node are in a 7-day time window (payouts occur 7-days after a post is made). Moreover, we divide the entire graph into three sub-graphs for train, validation, and test with \%80, \%10, and \%10 of heterogeneous graph data, respectively, which would be used as input into the GNN model, similar to \cite{3}.

\begin{table}[htbp]
  \centering
  \caption{Input features of all nodes}
  \label{features} 
  \resizebox{\columnwidth}{!}{%
    \begin{tabular}{lll}
    \toprule
    \textbf{Features} & \textbf{Node Type} & \textbf{Description} \\
    \midrule
    node\_type & post, user, comment & Type of node \\
    \midrule
    net\_rshares & post, comment & Sum of positive and negative rewards \\
    \midrule
    abs\_rshares & post, comment & Total absolute weight of votes \\
    \midrule
    vote\_rshares & post, comment & Total positive rshares from all votes \\
    \midrule
    author\_rewards & post, comment & Tracks the author payout  \\
    \midrule
    author\_reputation & post, comment & Author’s reputation \\
    \bottomrule
    \end{tabular}%
  }
\end{table}%

As we mentioned in Section~\ref{label:Heterogeneous Graph Construction}, we build the heterogeneous graph using NetworkX Python library, and for implementing GNN models we convert it into the form needed for the PyTorch Geometric (PyG) library. However, so far, there is no built-in function in the PyTorch library to convert a heterogeneous graph into PyG format. Therefore, this example also presents some insights to convert \emph{heterogeneous NetworkX graph} into a \emph{heterogeneous PyG graph}.

Heterogeneous graphs have different attributes within nodes and edges. Therefore, each node or edge has its own attributes with various dimensionality. As a result of heterogeneous data structure, standard Massage Passing (MP) in GNN cannot be easily applied. In the heterogeneous graph models, during message computation, the MP algorithm should run over edge dictionaries, and during node updates, it should iterate over node-type dictionaries.

\begin{figure}[htbp]
    \centering
    \includegraphics[scale=0.9]{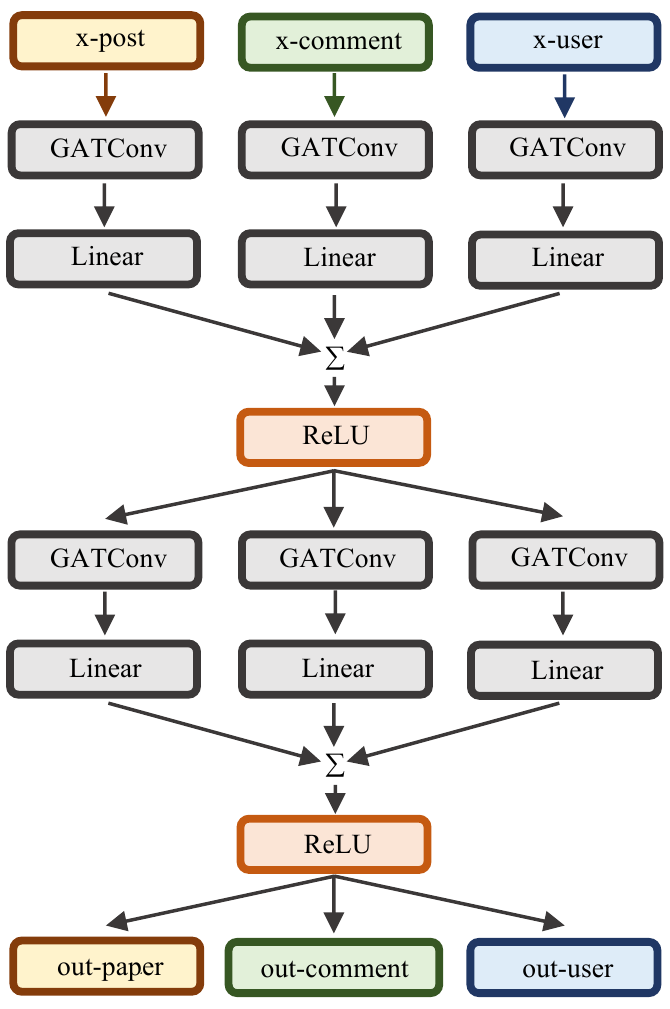}
    \caption{Model 1 for classification task}
    \label{GNN-model1}
\end{figure}

As shown in the Fig.~\ref{GNN-model1}, the method starts with a GAT-GNN model and repeats the message functions to operate separately on each edge type. As discussed earlier, we have three node types: user, post, and comment, that each has its own tensors data. At first, we normalized all tensors data and then entered them into the GAT convolution layer (each node type has its own layer) to aggregate temporal and topological features. The outputs of GAT layers are new sets of nodes features given to linear transformation to obtain sufficient expressive power of higher-level features. We then use the ReLU activation function to concatenate all the features and repeat this process one more time.

\subsubsection{Result}
We split the graph into three sub-graphs -- train contains 19200 posts, validation contains 2400 posts, and test contains 2400 posts. We trained the model using 128 hidden layers, and 5000 epochs with 0.001 learning rate. The model shows \%98.44, \%90.19, and \%87.44 accuracy for train, validation, and test sub-graphs, respectively. Table~\ref{model1-PC-1} shows performance criteria and confusion matrix of the model.

\begin{table}[htbp]
  \centering
  \caption{Model 1-3 Classes: Performance Criteria \& Confusion Matrix}
  \resizebox{\columnwidth}{!}{%
    \begin{tabular}{cccccccc}
    \toprule
    \multicolumn{4}{c}{\textbf{Performance Criteria}} &       & \multicolumn{3}{c}{\textbf{Confusion Matrix}} \\
    \toprule
          & Precision & Recall & F1-score &       & Low   & Medium & High \\
\cmidrule{1-4}\cmidrule{6-8}    Low   & 0.99  & 0.88     & 0.93  &       & 1964  & 277     & 0 \\
\cmidrule{1-4}\cmidrule{6-8}    Medium & 0.31     & 0.84     & 0.46     &       & 11   & 126     & 13 \\
\cmidrule{1-4}\cmidrule{6-8}    High  & 0.28     & 1.00     & 0.43     &       & 0     & 0     & 5 \\
    \bottomrule
    \end{tabular}%
    }
  \label{model1-PC-1}%
\end{table}%

\subsubsection{Sensitivity Analysis}
As we have an unbalanced dataset (i.e., non-zero payouts form \%30 of dataset), we decided to assign a separate class to zeros, and change the prediction problem from 3 classes into 4 classes to see whether the model performance changes. We use the same hyper-parameters as we had for 3 classes, and we got \%91.34, \%80.93, and \%79.38 accuracy for train, validation, and test sub-graphs, respectively. Table~\ref{model1-PC-2} shows performance criteria and confusion matrix for the model. The accuracy of the 4 class problem reduces from the 3 class problem. This indicates that the result of the 3 class problem was biased by unbalanced dataset. 

\begin{table}[htbp]
  \centering
  \caption{Model 1-4 Classes: Performance Criteria \& Confusion Matrix}
  \resizebox{\columnwidth}{!}{%
    \begin{tabular}{ccccccccc}
    \toprule
    \multicolumn{4}{c}{\textbf{Performance Criteria}} &       & \multicolumn{4}{c}{\textbf{Confusion Matrix}} \\
    \toprule
          & Precision & Recall & F1-score &       & Zeros & Low & Medium & High \\
\cmidrule{1-4}\cmidrule{6-9}    Zeros & 0.98  & 0.87     & 0.92  &       & 1575  & 235     & 0     & 0 \\
\cmidrule{1-4}\cmidrule{6-9}    Low & 0.45     & 0.45     & 0.45     &       & 17   & 195     & 219     & 0 \\
\cmidrule{1-4}\cmidrule{6-9}    Medium & 0.37     & 0.85     & 0.51     &       & 8   & 3     & 127     & 12 \\
\cmidrule{1-4}\cmidrule{6-9}    High & 0.29     & 1.00     & 0.45     &       & 0     & 0     & 0     & 5 \\
    \bottomrule
    \end{tabular}%
    }
  \label{model1-PC-2}%
\end{table}%

\subsection{Predicting Payout of a Post using Post Content}
In the second approach, we take the embedding vectors out of the body of the posts using a pre-trained Google Universal Sentence Encoder model. The model encodes text into 512-dimensional vectors ready to use for text classification, clustering, semantic similarity, etc. The pre-trained model is publicly available in Tensorflow-hub~\footnote{https://tfhub.dev/google/universal-sentence-encoder/4}. It is available in two variants, one trained using a Transformer encoder and the other with a Deep Averaging Network (DAN). There are trade-offs between these two variants' accuracy and computational resource requirements. The Transformer encoder is more accurate, despite its high computational cost, while the DNA encoder is less costly computational, despite its lower accuracy \cite{25}. To this end, we used the one with transformer to encodes the posts' content into 512-dimensional vectors, and gave them to the GNN model to predict new posts payout. 

\begin{figure}[htbp]
    \centering
    \includegraphics[scale=0.9]{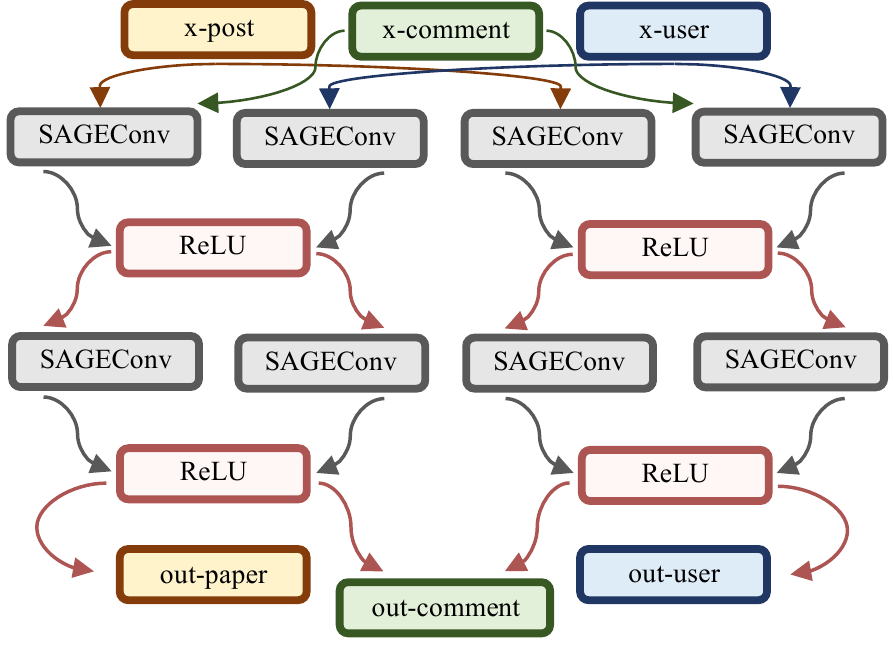}
    \caption{Model 2 for classification task}
    \label{GNN-model2}
\end{figure}

Again, as we have the heterogeneous graph, different types of node and edge features exist. Thus, the standard MP cannot easily applied and processed by the same function. As shown in the Fig.~\ref{GNN-model2}, the method starts with an SAGE-GNN model and repeats the message functions to operate separately on each edge type.

\subsubsection{Result}
We used the same split in training, validation and testing as described in Section~\ref{label:post-feature}. We trained the model using 512 hidden layers, and 3000 epochs with 0.001 learning rate. The model shows \%93.29, \%92.82, and \%93.53 accuracy for train, validation, and test sub-graphs, respectively.

As we have an unbalanced dataset (i.e., non-zero payouts form \%30 of dataset), we decided to assign a separate class to zeros, and change the prediction problem from 3 classes into 4 classes to see whether the model performance changes. We used the same hyper-parameters as we had for 3 classes. In this case, we got \%76.27, \%75.83, and \%75.54 accuracy for train, validation, and test sub-graphs, respectively. This indicates that the result of the 3 class problem was biased by unbalanced dataset.

\section{Conclusion}
In this paper, we presented a new approach to representing social networks as dynamic heterogeneous graphs. We showed how this representation handles the inherent multiplexity that is part of real-world social networks, while also representing time explicitly in a way that enables powerful querying and predictive tasks. We presented several examples using a real-world heterogeneous social network, Steemit, where there are monetary incentives for users as well. We demonstrated how common queries on the Steemit social media graph can be executed efficiently using this graph structure. We also demonstrate how deep learning based predictive analytics can be applied using this graph structure on the Steemit data. As far as we know this is one of the first papers to have modeled real-world social networks using both time and heterogeneous graph representations. 

There are many applications as well as important opportunities for future work. It is well-known that representation - and how data is stored in the back-end of a system - drives functionalities that can be provided on the front-end to users. Representations such as the one presented here can more easily enable a range of new user-functionalities in social media platforms that can allow users to query items (e.g. other users, or content types such as posts, videos, comments etc.) based on temporal constraints. 

Further, if deep learning on graphs extends to dynamic heterogeneous graphs it can allow many new predictive models to also be developed on such data. These models may for example, in real-time, be able to predict how much engagement a specific user's post or comment may receive in a short span of future time. Such predictions can help platforms improve ranking algorithms, and can also be used to offer users suggestions on how to develop content that can be more engaging. This is important in platforms such as TikTok where there is an entire ecosystem of content creators who depend on engagement of content for their livelihoods. 

However, while our work here offers a potentially more expressive representation of social media data, there are challenges that need to be addressed in future work to make the functionalities mentioned previously practical. These include work on query optimization that can scale queries on dynamic heterogeneous graphs such that these can be executed in real-time. Also, there are many opportunities to design better deep learning algorithms specifically tailored for dynamic heterogeneous graphs.

\end{document}